\documentclass[aps,prl,preprint,superscriptaddress,floatfix]{revtex4-1}
\usepackage{graphicx}
\usepackage{amssymb}
\usepackage{amsfonts}
\usepackage{amsmath}

\begin{document}

\title{Rotational symmetry breaking in the topological superconductor Sr$_x$Bi$_2$Se$_3$ probed by upper-critical field experiments}

\author{Y. Pan} \email{y.pan@uva.nl} \affiliation{Van der Waals - Zeeman Institute, University of Amsterdam, Science Park 904, 1098 XH Amsterdam, The Netherlands}
\author{A. M. Nikitin}\affiliation{Van der Waals - Zeeman Institute, University of Amsterdam, Science Park 904, 1098 XH Amsterdam, The Netherlands}
\author{G. K. Araizi}\affiliation{Van der Waals - Zeeman Institute, University of Amsterdam, Science Park 904, 1098 XH Amsterdam, The Netherlands}
\author{Y. K. Huang} \affiliation{Van der Waals - Zeeman Institute, University of Amsterdam, Science Park 904, 1098 XH Amsterdam, The Netherlands}
\author{Y. Matsushita} \affiliation{National Institute for Materials Science, Sengen 1-2-1, Tsukuba, Ibaraki 305-0047, Japan}
\author{T. Naka} \affiliation{National Institute for Materials Science, Sengen 1-2-1, Tsukuba, Ibaraki 305-0047, Japan}
\author{A. de Visser} \email{a.devisser@uva.nl} \affiliation{Van der Waals - Zeeman Institute, University of Amsterdam, Science Park 904, 1098 XH Amsterdam, The Netherlands}

\date{\today}

\begin{abstract}
Recently it was demonstrated that Sr intercalation provides a new route to induce superconductivity in the topological insulator Bi$_2$Se$_3$. Topological superconductors are predicted to be unconventional, with mixed even and odd parity Cooper pairs states. An adequate probe to test for unconventional superconductivity is the upper critical field, $B_{c2}$. For a standard BCS layered superconductor $B_{c2}$ shows an anisotropy when the magnetic field is applied parallel and perpendicular to the layers, but is isotropic when the field is rotated in the plane of the layers. Here we report measurements of the upper critical field of superconducting Sr$_x$Bi$_2$Se$_3$ crystals ($T_c = 3.0$~K). Surprisingly, field-angle dependent magnetotransport measurements reveal a large anisotropy of $B_{c2}$ when the magnet field is rotated in the basal plane. The large two-fold anisotropy, while six-fold is anticipated, cannot be explained with the Ginzburg-Landau anisotropic effective mass model or flux flow induced by the Lorentz force. The rotational symmetry breaking of $B_{c2}$ indicates unconventional superconductivity with an odd-parity polarized triplet Cooper pair state ($\Delta _4$-pairing) recently proposed for rhombohedral topological superconductors, or might have a structural nature, such as self-organized stripe ordering of Sr atoms.
\end{abstract}

\date{\today}


\maketitle

Currently, topological insulators (TIs) are at the focus of condensed matter research, because they offer  unprecedented possibilities to study novel quantum states~\cite{Hasan&Kane2010,Qi&Zhang2010,Ando2013}. 3D TIs are bulk insulators with a non-trivial topology of the electron bands that gives rise to surface states at the edge of the material. The gapless surface states have a Dirac-type energy dispersion with the spin locked to the momentum and are protected by symmetry. This makes TIs promising materials for applications in fields like spintronics and magnetoelectrics~\cite{Hasan&Kane2010,Qi&Zhang2010}. The concept of a TI can also be applied to superconductors (SCs), where the superconducting gap corresponds to the gap of the band insulator~\cite{Kitaev2009,Schnyder2009}. Topological superconductors are predicted to be unconventional, with a mixed even and odd parity Cooper pair state~\cite{Sato2009,Sato2010}. Much research efforts are devoted to 1D and 2D SCs, where Majorana zero modes exist as protected states at the edge of the SC~\cite{Mourik2012,Beenakker2013}. Majorana zero modes with their non-Abelian statistics offer a unique platform for future topological quantum computation devices~\cite{Nayak2008}. Prominent candidates for 3D topological SC are the Cu intercalated TI Bi$_2$Se$_3$~\cite{Hor2010,Kriener2011a}, the doped topological crystalline insulator Sn$_{1-x}$In$_{x}$Te~\cite{Sasaki2012} and selected topological half-Heusler compounds~\cite{Butch2011,Yan&deVisser2014,Nakajima2015}.

Among the 3D topological superconductors, Cu$_x$Bi$_2$Se$_3$, which has a SC transition temperature $T_c = 3$~K for $x=0.3$~\cite{Hor2010,Kriener2011a}, is the most intensively studied material. ARPES (Angle Resolved PhotoEmission Spectroscopy) experiments conducted to study the bulk and surface states reveal that the topological character is preserved when Bi$_2$Se$_3$ is intercalated with Cu~\cite{Wray2010}. By evaluating the topological invariants of the Fermi surface, Cu$_x$Bi$_2$Se$_3$ is expected to be a time-reversal invariant fully-gapped odd-parity topological SC~\cite{Sato2009,Sato2010}. This was put on a firmer footing by a two-orbital pairing potential model where odd-parity SC is favoured by strong spin-orbit coupling~\cite{Fu&Berg2010}. Several experiments have been interpreted in line with topological SC. The specific heat shows a full SC gap~\cite{Kriener2011a}. The upper critical field exceeds the Pauli limit and has a temperature variation that points to spin-triplet SC~\cite{Bay2012}. Much excitement was generated by the observation of a zero-bias conductance peak in point contact spectroscopy, that was attributed to a Majorana surface state~\cite{Sasaki2011}. However, STS (Scanning Tunneling Spectroscopy) showed that the density of states at the Fermi level is fully gapped without any in-gap states~\cite{Levy2013}. On the other hand, the superconducting state shows a large inhomogeneity~\cite{Levy2013} and the superconducting volume fraction depends on quenching conditions~\cite{Schneeloch2015}. Consequently, the issue of topological SC in Cu$_x$Bi$_2$Se$_3$ has not been settled and further experiments are required, as well as new materials.

Very recently it has been demonstrated that Sr intercalation provides a new route to induce superconductivity in Bi$_2$Se$_3$~\cite{Liu2015}. Resistivity and magnetization measurements on Sr$_x$Bi$_2$Se$_3$ single crystals with $x=0.06$ show $T_c = 2.5$~K. The SC volume fraction amounts to 90~\% which confirms bulk SC. By optimizing the Sr content a maximum $T_c$ of 2.9 K was found for $x=0.10$~\cite{Shruti2015}. The topological character of Bi$_2$Se$_3$ is preserved upon Sr intercalation. ARPES showed a topological surface state well separated from the bulk conduction band~\cite{Han2015,Neupane2015}. Based on the first measurements of the electronic parameters in the normal and SC states, and the close analogy to Cu$_x$Bi$_2$Se$_3$, it has been advocated that Sr$_x$Bi$_2$Se$_3$ is a new laboratory tool to investigate topological SC~\cite{Liu2015,Shruti2015}.

Here we report a study of unusual basal-plane anisotropy effects in the upper critical field, $B_{c2}$, of Sr$_x$Bi$_2$Se$_3$. Bi$_2$Se$_3$ crystallizes in a rhombohedral structure with space group $R \overline{3}m$. It is a layered material and Sr is intercalated in the Van der Waals gaps between the quintuple Bi$_2$Se$_3$ layers~\cite{Liu2015}. For a standard BCS (Bardeen, Cooper, Schrieffer) layered SC the anisotropy of $B_{c2}$ is expressed by the parameter $\gamma ^{an} = B_{c2}^{\parallel}/B_{c2}^{\perp}$, where $B_{c2}^{\parallel}$ and $B_{c2}^{\perp}$ are measured with the $B$-field parallel and perpendicular to the layers, respectively~\cite{Klemm2012}. Whereas $B_{c2}^{\parallel}$ is normally isotropic, Sr$_x$Bi$_2$Se$_3$ presents a unique exception. Field-angle-dependent magnetotransport experiments demonstrate a large two-fold basal-plane anisotropy of $B_{c2}$, with $B_{c2}^{a} = 7.4$~T and $B_{c2}^{a^*} = 2.3$~T for $x=0.15$ at $T/T_c = 0.1$ ($T_c = 3.0$~K), where $a$ and $a^*$ are orthogonal directions in the basal plane. This large effect cannot be explained with the anisotropic effective mass model~\cite{Morris1972,Klemm2012} or the variation of $B_{c2}$ caused by flux flow~\cite{Tinkham1996}. The rotational symmetry breaking of $B_{c2}$ indicates unconventional superconductivity~\cite{Nagai2012,Fu2014}, or might have a structural nature, such as preferential ordering of Sr atoms.

\vspace{5mm}
\noindent
\textbf{Results}

\noindent
The resistivity, $\rho (T)$, of our Sr$_{x}$Bi$_{2}$Se$_3$ crystals with $x=0.10$ and $x=0.15$ shows a metallic temperature variation with superconducting transition temperatures $T_c$ of 2.8~K and 3.0~K, respectively, see Fig.~S4 in the Supplementary Information~\cite{Pan-SI2016}. The SC volume fractions of the crystals measured by ac-susceptibility amount to 40~\% and 80~\%, respectively~\cite{Pan-SI2016}. In Fig.~1 we show the angular variation of the resistance, $R(\theta)$, measured in a fixed field $B=0.4$~T directed in the basal plane ($aa^*$-plane), in the temperature range 2-3~K around $T_c$ ($T_c =2.8~$K at $B=0$~T), for $x=0.10$. Rather than attaining a constant value, the curves show a pronounced angular variation which demonstrates that $B_{c2}(T)$ (or $T_c(B)$) is field-angle dependent. For instance, at 2.5~K and 0.4~T (violet symbols) the sample is in the normal state at $\theta = 3^\circ$ and superconducts ($R=0$) at $93^\circ$. By raising the temperature from 2~K to 3~K SC is smoothly depressed for all field directions. The data show a striking two-fold symmetry, which is most clearly demonstrated in a polar plot (Figure~2). We remark, the same two-fold anisotropy is observed in crystals with $x=0.15$. In the top panel of Fig.~1 we show $R(\theta)$ in the normal state measured in 8~T for $x=0.10$. The data have been symmetrized after measuring $R(\theta)$ for opposite field polarities to eliminate a small Hall component. $R(\theta)$ in the normal state shows the same two-fold symmetry as in Fig.~1a. The variation in $R(\theta)$ is small and amounts to 3~\% in 8~T. The data follow a $\sin \theta$ dependence, which tells us the variation is due to the classical magnetoresistance related to the Lorentz force $F_L = BI\sin \theta$, where $I$ is the transport current that flows in the basal plane. $R(\theta)$ is minimum in the longitudinal case ($B \parallel I$) and maximum in the transverse case ($B \perp I$).

In Fig.~3 we report $B_{c2}(T)$ for two single crystals measured with the $B$-field along the orthogonal directions in the hexagonal unit cell. The data points are obtained by measuring the superconducting transition in $R(T)$ in fixed fields, where $T_c$ is identified by the 50~\% drop of $R$ with respect to its value in the normal state~\cite{Pan-SI2016}. In determining the values of $B_{c2}$ we did not correct for demagnetization effects, since the demagnetization factors calculated for our crystals are small~\cite{Pan-SI2016}. As expected from the data in Fig.~1, we observe a large difference between $B_{c2}^a$ and $B_{c2}^{a^*}$, with an in-plane anisotropy parameter $\gamma _{aa^*}^{an} = B_{c2}^a / B_{c2}^{a^*}$ of 6.8 (at 1.9~K) and 2.6 (at 0.3~K) for $x=0.10$ and $x=0.15$, respectively. For both crystals $B_{c2}^{a^*} \approx B_{c2}^c$. Obviously, the $B_{c2}$ ratio $\gamma ^{an}$ for the field $\parallel$ and $\perp$ to the layers depends on the field angle and ranges from 1.2 to 3.2 for $x=0.15$. In Ref.~\onlinecite{Shruti2015} a value for $\gamma ^{an}$ of 1.5 is reported, whereas from the data in Ref.~\onlinecite{Liu2015} we infer a value of 1. In the top panels of Fig.~3 we show $\rho (B)$ measured along the $a$, $a^*$ and $c$ axis at $T=2.0$~K and $T=0.3$~K for $x=0.10$ and $x=0.15$, respectively. The $B_{c2}(T)$ values are determined by the midpoints of the transitions to the normal state, and are indicated by open symbols in the lower panels. The agreement between both methods (field sweeps and temperature sweeps) is excellent. For the $x=0.15$ sample we see a remarkable broadening for $B \parallel a$. The initial small increase of $\rho (B)$ between 4 and 6~T is most likely related to a sample inhomogeneity, because a similar tail is also observed in the $R(T)$ data~\cite{Pan-SI2016}.

In Fig.~4 we show the angular variation of the upper critical field, $B_{c2}(\theta)$. For this experiment the crystals are placed on the rotator and the field is oriented in the basal plane. The data points are obtained as the midpoints of the transitions to the normal state of the $R(B)$ curves measured at temperatures of 2~K for $x=0.10$ and of 0.3~K and 2~K for $x=0.15$ (see Fig.~S7~\cite{Pan-SI2016}). All data sets show the pronounced two-fold basal-plane anisotropy of $B_{c2}$, already inferred from Figs.~1 and 2.

\vspace{5mm}
\noindent
\textbf{Discussion}

\noindent
Having conclusively established the two-fold anisotropy of $B_{c2}$ in the basal plane, we now turn to possible explanations. A first explanation could be a lowering of the symmetry caused by a crystallographic phase transition below room temperature. However, the powder X-ray diffraction patterns measured at room temperature and $T=10$~K are identical (see Fig.~S2 in Ref.~\onlinecite{Pan-SI2016}). Moreover, the resistivity traces ($T=2-300$~K, Fig.~S8) and the specific heat ($T=2-200$~K, Fig.~S6) all show a smooth variation with temperature and do not show any sign of a structural phase transition~\cite{Pan-SI2016}. We therefore argue our crystals keep the $R \overline{3}m$ space group at low temperatures.

A second explanation for breaking the symmetry in the basal plane could be the measuring current itself. Since the current flows in the basal plane it naturally breaks the symmetry when we rotate the field in the basal-plane. Indeed $B_{c2}$ is largest for $B \parallel I$ and smallest for $B \perp I$. In the latter geometry, and for large current densities, the Lorentz force may cause flux lines to detach from the pinning centers, which will lead to a finite resistance, a broadened $R(B)$-curve and a lower value of $B_{c2}$~\cite{Tinkham1996}. This effect has been observed for instance in the hexagonal superconductor MgB$_{2}$ by rotating $B$ with respect to $I$ in the basal plane~\cite{Shi2003}. For a current density 30 A/cm$^2$, the two-fold anisotropy obtained just below $T_c = 36$~K is small, $\sim$~8~\%~\cite{Shi2003}. In our transport experiments the current densities are $\leq 0.4 $~A/cm$^2$ and we did not detect a significant effect on the resistance when the current density was varied close to $T_c$~(see Fig.~S9 \cite{Pan-SI2016}). Also, when flux flow has a significant contribution, one expects the $R(B)$-curves for $B \perp I$ to be broader than the curves for $B \parallel I$. However, we observe the reverse (see Figs~3a,b). Moreover, the anisotropy is still present at $T/T_c = 0.1$ and is much larger (of the order of 300~\%, see Fig.~4) than can be expected on the basis of flux flow. In order to further rule out the influence of the current direction we have investigated $B_{c2} (\theta)$ in the basal plane with the transport current perpendicular to the layers ($I \parallel c$) and thus keeping $B \perp I$ (see Fig.~S11, Ref.~\onlinecite{Pan-SI2016}). The angular variation of the resistance, measured in this geometry using a two-probe method, is similar to that reported in Fig.~1. Thus the two-fold anisotropy in $B_{c2}$ is also present for the $B$-field in the $aa^*$-plane and the current along the $c$-axis.

Next we address whether the variation of $B_{c2}$ in the basal plane can be attributed to the anisotropy of the effective mass. Within the Ginzburg-Landau model~\cite{Takanaka1982,Klemm2012} the anisotropy of $B_{c2}$ is attributed to the anisotropy of the SC coherence length, $\xi$, which in turn relates to the anisotropy of the effective mass. For a layered superconductor the anisotropy ratio $\gamma ^{an} = B_{c2}^{\parallel}/B_{c2}^{\perp} = \sqrt{M/m}$~\cite{Morris1972}. Here $m$  and $M$ are the effective masses $\parallel$ and $\perp$ to the layers. In the rhombohedral structure $m =m_a = m_{a^*}$ and $M = m_c$, where the subscripts $a$, $a^*$ and $c$ refer to the effective masses for the energy dispersion along the main orthogonal crystal axes (\textit{i.e.} in the hexagonal unit cell). For a field rotation in the $aa{^*}$-plane $B_{c2}^{\parallel}$ is in general isotropic, since $m_{a} \approx  m_{a{^*}} (< m_c)$. For a 3D anisotropic superconductor the angular variation $B_{c2}(\theta)$ in a principal crystal plane can be expressed as $B_{c2}(\theta) = B_{c2} (0^{\circ})/(\cos^2 \theta + \gamma^{-2} \sin^2 \theta )^{1/2}$, where $\gamma = B_{c2}(90^{\circ})/B_{c2}(0^{\circ})$. To provide an estimate of $\gamma$ for Sr$_{0.15}$Bi$_{2}$Se$_3$, we compare in Fig.~4b the measured $B_{c2}(\theta)$ with the angular variation in the anisotropic effective mass model (solid line). We obtain $B_{c2}(0^{\circ})= 2.3$~T, $B_{c2}(90^{\circ})= 7.4$~T and $\gamma = 3.2$. The effective mass ratio $m_{a^*}/m_a = \gamma^2$~\cite{Takanaka1982} would then attain the large value of 10.2. As we show below, this is not compatible with the experimental Fermi-surface determination.

The Fermi surface of $n$-doped Bi$_2$Se$_3$, with a typical carrier concentration $n \sim  2\times10^{19}$~cm$^{-3}$ representative for the SC Sr$_{x}$Bi$_{2}$Se$_3$ crystals~\cite{Liu2015,Shruti2015}, has been investigated by the Shubnikov - de Haas effect~\cite{Koehler1973,Lahoud2013,Liu2015}. It can be approximated by an ellipsoid of revolution with the longer axis along the $k_c$-axis. A trigonal warping of the Fermi surface due to the rhombohedral symmetry has been detected, but the effect is small: the variation of the effective mass in the basal plane amounts to a few \% only ~\cite{Koehler1973}. This also explains why $R(\theta)$ in the normal state (Fig.~1a), does not show a $2 \pi /3$ periodicity superimposed on the two-fold symmetry induced by the current. Clearly, the two-fold symmetry (Fig.~4), while three fold is expected, and the calculated large ratio $m_{a^*}/m_a$ using the Ginzburg-Landau model are at variance with the experimental Fermi-surface determination~\cite{Koehler1973} and we discard this scenario.

We remark that the overall temperature variation $B_{c2}(T)$ reported in Fig.~3d is at variance with the standard BCS behaviour for a weak-coupling spin-singlet SC~\cite{Werthamer1966}. All the curves show an upward curvature below $T_c$, followed by a quasi-linear behaviour down to the lowest temperatures. Furthermore, for $B \parallel a$, $B_{c2}(T \rightarrow 0)$ largely exceeds the Pauli limit $B^P (T \rightarrow 0) = 1.86 \times T_c \approx 5.6$~T for a spin-singlet SC~\cite{Clogston1962}. This may point to an odd-parity component in the SC order parameter. Nagai~(Ref.~\onlinecite{Nagai2012}) and Fu~(Ref.~\onlinecite{Fu2014}) recently proposed a model for odd parity polarized spin-triplet SC developed in the context of Cu$_x$Bi$_2$Se$_3$, and investigated the experimental consequences of $\Delta _4$ pairing in the two-orbital model~\cite{Fu&Berg2010}. Here, SC is described by an odd-parity two-dimensional representation, $E_u$, where the attractive potential pairs two electrons in the unit cell to form a spin triplet, \textit{i.e.} a linear combination of $c_{1\uparrow}c_{2\uparrow}$ and $c_{1\downarrow}c_{2\downarrow}$. The indices 1,2 refer to the two orbitals and the arrows to the spin. The $\Delta _4$ state has zero-total spin along an in-plane direction $\textbf{n}=(n_x,n_y)$ that is regarded as a nematic director and breaks rotational symmetry. By taking into account the full crystalline anisotropy in the Ginzburg-Landau model, it can be shown that $\mathbf{n}$ is pinned to a direction in the basal plane. For $\mathbf{n}= \mathbf{\hat{x}}$, point nodes in the SC gap are found along $\mathbf{\hat{y}}$, whereas for $\mathbf{n}= \mathbf{\hat{y}}$ two gap minima occur at $\pm k_F \mathbf{\hat{x}}$~\cite{Fu2014}. Our $B_{c2}$-data can be interpreted as reflecting a strongly anisotropic SC gap function. The SC coherence length, $\xi$, along the main axes can be evaluated from the Ginzburg-Landau relations $B_{c2}^a = \Phi _0 / (2 \pi \xi_{a^*} \xi_c )$, $B_{c2}^{a^*} = \Phi _0 / (2 \pi \xi_{a} \xi_c )$ and $B_{c2}^c = \Phi _0 / (2 \pi \xi_{a} \xi_{a^*})$. Here $\Phi _0$ is the flux quantum. With the experimental $B_{c2}$-values, taken at $T/T_c = 0.1$ in Fig.~3d for $x=0.15$, we calculate $\xi _a = 19.6$~nm, $\xi _{a^*} = 7.6$~nm and $\xi _c = 5.4$~nm. Interpreting $\xi$ as the Cooper-pair size, this implies that the pairing interaction is strongest along the $a^*$ and $c$-axis, and weakest along the $a$-axis. The observation that $\xi_a > \xi_{a^*} \approx \xi_c$, is in agreement with a gap structure with $\mathbf{n}= \mathbf{\hat{y}}$. On the other hand, in the $\Delta _4$ pairing model rotational symmetry breaking due to nematic order is a property of the SC state, while $B_{c2}$ probes the transition to the normal state and therefore should retain the hexagonal symmetry of the crystal lattice~\cite{Fu2015,Krotkov2002}. This indicates a more intricate scenario. We remark that rotational symmetry breaking in the spin system has been observed by Nuclear Magnetic Resonance (NMR) in the related superconductor Cu$_x$Bi$_2$Se$_3$, which is considered to provide solid evidence for a spin-triplet state~\cite{Matano2015}.

Yet another interesting possibility is a self-organized structural stripiness in the optimum for superconductivity due to ordering of Sr atoms in the Van der Waals gaps. This could naturally lead to an anisotropy of $B_{c2}$ when measured for a current in the basal plane, because of an effective reduced dimensionality. The higher $B_{c2}$-values will then be found for $B \parallel I$ along the stripes. On the other hand, for $I$ perpendicular to the layers the basal-plane anisotropy of $B_{c2}$ is found as well~\cite{Pan-SI2016}. This calls for a detailed compositional and structural characterization of Sr$_x$Bi$_2$Se$_3$ by techniques such as Electron Probe Microprobe Analysis (EPMA) or Transmission Electron Microscopy (TEM). Notice that in Cu$_x$Bi$_2$Se$_3$ crystals EPMA has revealed that the Cu concentration shows variations on the sub-mm scale, which gives rise to SC islands~\cite{Kriener2011b}. Moreover, a STM study reports an oscillatory behaviour of the Cu pair distribution function due to screened Coulomb repulsion of the intercalant atoms~\cite{Mann2014}.

In conclusion, we have investigated the angular variation of the upper critical field of superconducting crystals of Sr$_{x}$Bi$_{2}$Se$_3$. The measurements reveal a striking two-fold anisotropy of the basal-plane $B_{c2}$. The large anisotropy cannot be explained with the anisotropic effective mass model or the variation of $B_{c2}$ caused by flux flow. We have addressed two alternative explanations: (i) unconventional superconductivity, with an odd-parity triplet Cooper-pair state ($\Delta _4$ pairing), and (ii) self-organized striped superconductivity due to preferential ordering of Sr atoms. The present experiments and results provide an important benchmark for further unravelling the superconducting properties  of the new candidate topological superconductor Sr$_{x}$Bi$_{2}$Se$_3$.

\vspace{5mm}
\noindent
\textbf{Methods}

\noindent
\textbf{Sample preparation.} Single crystals Sr$_x$Bi$_2$Se$_3$ with $x=0.10$ and $x=0.15$ were prepared by melting high-purity elements at 850~$^\circ$C in sealed evacuated quartz tubes, followed by slowly cooling till 650~$^{\circ}$C at the rate of 3~$^\circ$C/hour. Powder X-ray diffraction confirms the $R \overline{3}m$ space group (see Supplementary Information~\cite{Pan-SI2016}). Laue back-scattering diffraction confirmed the single-crystallinity and served to identify the crystal axes $a$ and $a^*$. Thin bar-like samples with typical dimensions $0.3 \times 1.5 \times 3$ mm$^3$ were cut from the bulk crystal for the transport measurements.

\vspace{5mm}
\noindent
\textbf{Magnetotransport experiment.}
Magnetotransport experiments were carried out in a PPMS-Dynacool (Quantum Design) in the temperature range from 2~K to 300~K and magnetic fields up to~9 T and in a 3-Helium cryostat (Heliox, Oxford Instruments) down to 0.3~K and fields up to 12~T. The resistance was measured with a low-frequency ac-technique in a 4-point configuration with small excitation currents, $I$, to prevent Joule heating ($I$ = 0.5-1~mA in the PPMS and 100~$\mu$A in the Heliox experiments). The current was applied in the basal plane along the long direction of the sample. For in-situ measurements of the angular magnetoresistance the crystals were mounted on a mechanical rotator in the PPMS and a piezocrystal-based rotator (Attocube) in the Heliox. The samples were mounted such that the rotation angle $\theta    \simeq 0^\circ$  corresponds to $B \perp I$. Care was taken to align the $a$-axis with the current direction, but a misorientation of several degrees can not be excluded.

\vspace{5mm}
\noindent
\textbf{Acknowlegdments}

\noindent
The authors acknowledge discussions with A. Brinkman, U. Zeitler, R.J. Wijngaarden and Liang Fu. This work was part of the research program on Topological Insulators funded by FOM (Dutch Foundation for Fundamental Research of Matter).

\vspace{5mm}
\noindent
\textbf{Author contributions}
\noindent
Y.P.: magnetotransport and ac-susceptibility in the PPMS, data analysis; A.M.N. and G.K.A.: magnetotransport in the Heliox. H.Y.K.: crystal synthesis and Laue single-crystal diffraction; Y.M. and T.N.: temperature dependent X-ray measurements. A.d.V.: experiment design, supervision measurements, manuscript writing with contributions of Y.P.

\vspace{5mm}
\noindent
\textbf{Additional Information}
\noindent
Supplementary Information accompanies this paper.

\newpage

\bibliography{Refs_SrBi2Se3_sci_rep}

\newpage
\begin{figure}
\includegraphics[width=12cm]{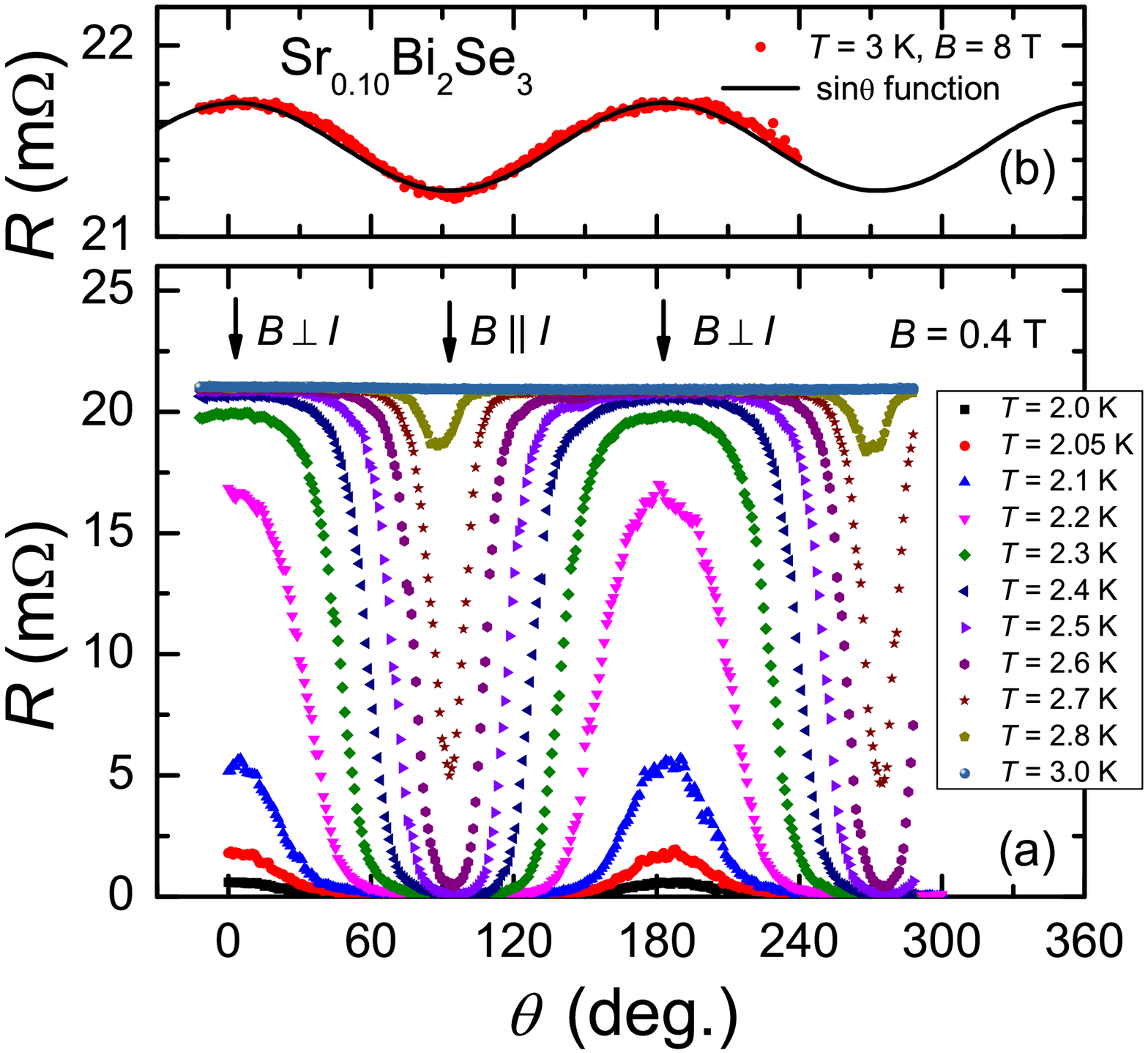}
\caption{\textbf{Angular variation of the resistance of Sr$_{0.10}$Bi$_{2}$Se$_3$.} Lower panel: Resistance of Sr$_{0.10}$Bi$_{2}$Se$_3$ as a function of angle $\theta$ at $B= 0.4$~T and temperatures between 2.0~K (bottom) and 3.0~K (top). The angle $\theta = 3^\circ$ corresponds to $B \perp I$ and $\theta = 93^\circ$ to $B \parallel I$ as indicated by arrows. The current direction is along the $a$-axis, with a precision of several degrees. The data are measured with increasing angle, and reproduce when the rotation direction is reversed, apart from a small backlash in the rotator of $2 ^\circ$. Upper panel: $R(\theta)$ in the normal state at $T=3.0$~K and $B=8$~T. The solid line shows $R( \theta)$ can be described by a $\sin \theta$ function.}
\end{figure}

\begin{figure}
\includegraphics[width=12cm]{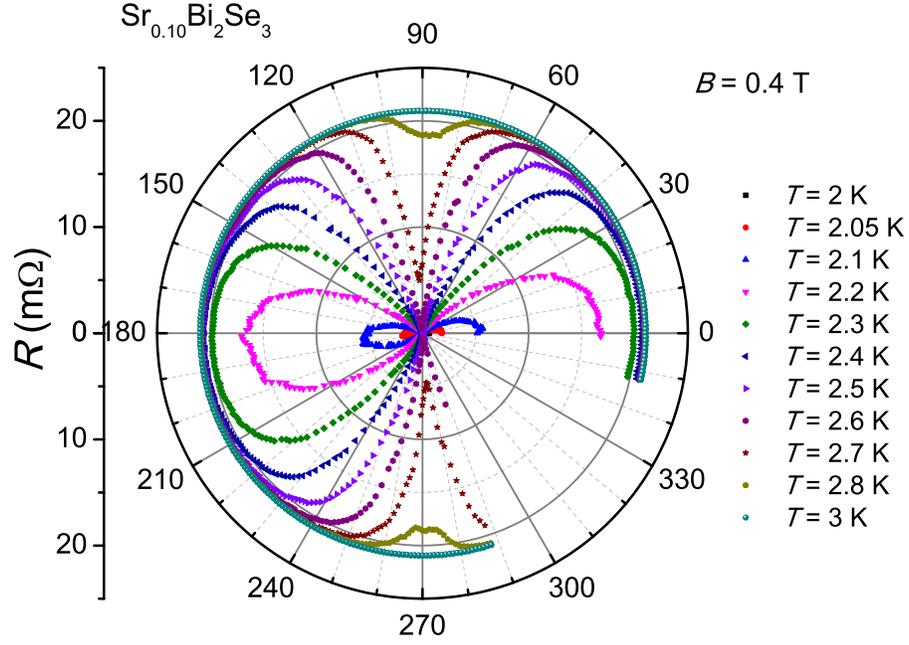}
\caption{\textbf{Polar plot of the resistance of Sr$_{0.10}$Bi$_{2}$Se$_3$.} Resistance of Sr$_{0.10}$Bi$_{2}$Se$_3$ as a function of angle $\theta$ in a magnetic field of 0.4~T and temperatures ranging from 2.0~K to 3.0~K presented in a polar plot. The angle $\theta = 3^\circ$ corresponds to $B \parallel a^* \perp I$, while $\theta = 93^\circ$ corresponds to $B \parallel a \parallel I$.}
\end{figure}

\begin{figure}
\includegraphics[width=12cm]{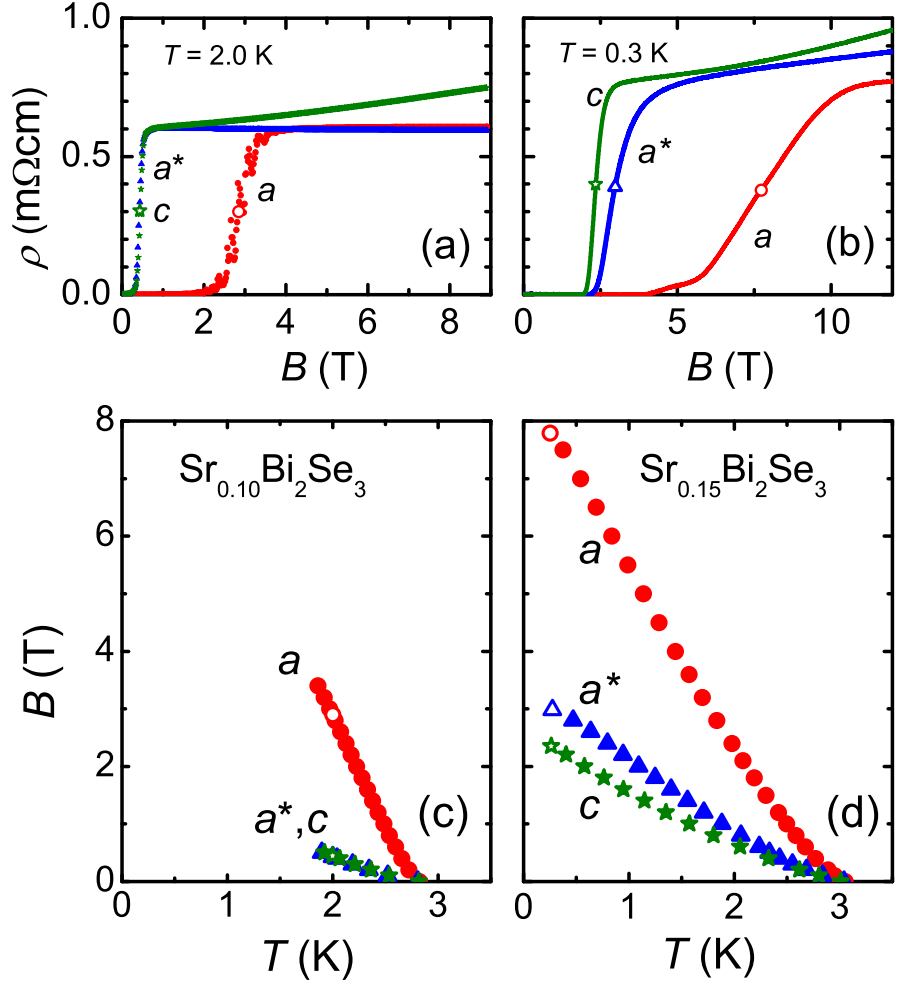}
\caption{\textbf{Upper critical field of Sr$_x$Bi$_2$Se$_3$.} Panel (a) and (b): Resistance of Sr$_{x}$Bi$_{2}$Se$_3$ as a function of $B \parallel$ $a$, $a^*$ and $c$, for $x=0.10$ and 0.15, respectively. The open symbols indicate the midpoints of the transitions to the normal state. Panel (c) and (d): $B_{c2}$ obtained for $B \parallel$ $a$, $a^*$ and $c$, for $x=0.10$ and 0.15, respectively. Solid symbols from midpoints of $R(T)$-curves in fixed $B$~\cite{Pan-SI2016}. Open symbols from $\rho (B)$ at fixed $T$. In the experiments for $x=0.15$ the crystal was not mounted on the rotator but oriented by eye, which adds some inaccuracy as regards field alignment. The current direction was always along the $a$-axis, with a precision of several degrees.}
\end{figure}

\begin{figure}
\includegraphics[width=12cm]{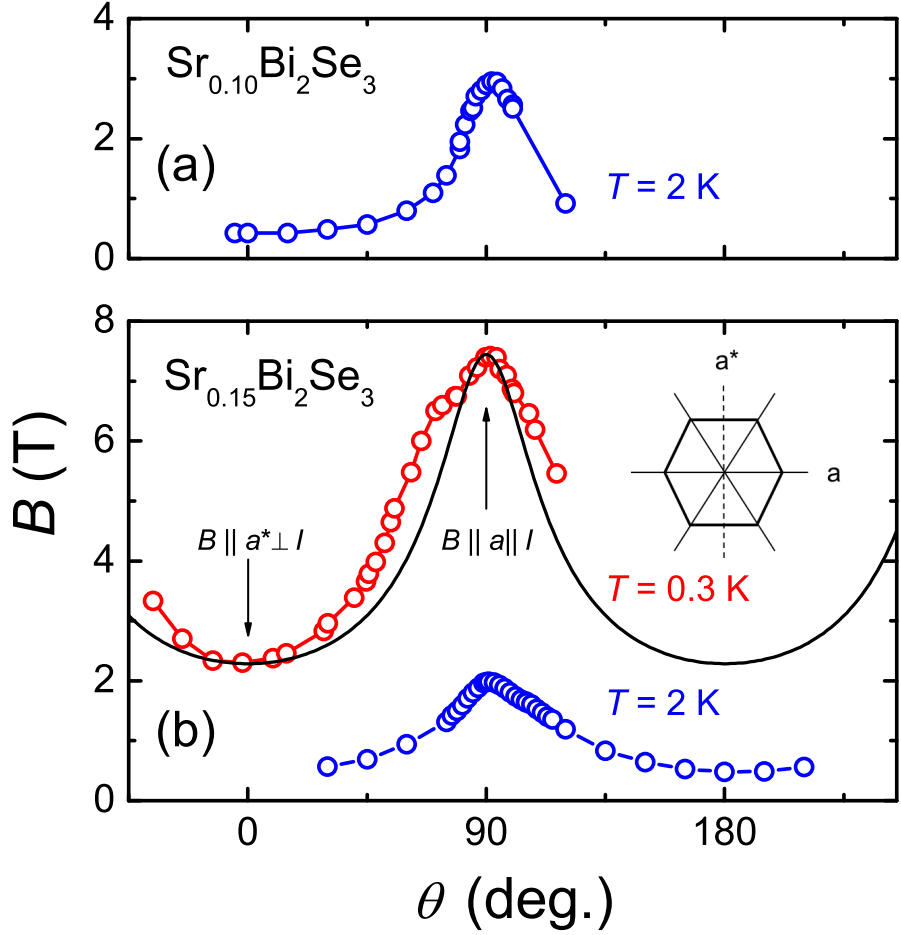}
\caption{\textbf{Angular variation of $B_{c2}$ of Sr$_x$Bi$_2$Se$_3$ in the basal plane.} Panel (a) and (b): Angular variation of $B_{c2}$ in the basal plane ($aa^*$-plane) for $x=0.10$ and 0.15, respectively, at temperatures as indicated. The data are obtained from $R(B)$ measurements at fixed $T$. The angle $\theta = 0^\circ$ corresponds to $B \parallel a^* \perp I$ and $\theta = 90^\circ$ to $B \parallel a \parallel I$. The solid black line in panel (b) represents $B_{c2} (\theta)$ for an anisotropic effective mass model with two-fold symmetry and $\gamma = 3.2$ (see text). The $a$ and $a^*$ directions in the hexagonal basal plane are defined as in the figure in the upper right corner of panel (b).}
\end{figure}

\end{document}